# Ferroelectric Rashba Semiconductors as a novel class of multifunctional materials

Silvia Picozzi[1]

[1]Consiglio Nazionale delle Ricerche – Institute for Superconducting and Innovative materials and devices (CNR–SPIN, L'Aquila)

\* **Correspondence:** Silvia Picozzi, CNR– SPIN L'Aquila, Via Vetoio 10, 67100 L'Aquila, ITALY, silvia.picozzi@spin.cnr.it

Keywords: Ferroelectricity, Rashba effect, spin-orbit coupling, electronic structure theory, GeTe, semiconductor spintronics.

## Abstract

The discovery of novel properties, effects or microscopic mechanisms in modern materials science is often driven by the quest for combining, into a single compound, several functionalities: not only the juxtaposition of the latter functionalities, but especially their coupling, can open new horizons in basic condensed matter physics, in materials science and technology. Semiconductor spintronics makes no exception. In this context, we have discovered by means of density-functional simulations that, when a sizeable spin-orbit coupling is combined with ferroelectricity, such as in GeTe, one obtains novel multifunctional materials - called Ferro-Electric Rashba Semi-Conductors (FERSC) - where, thanks to a giant Rashba spin-splitting, the spin texture is controllable and switchable via an electric field. This peculiar spin-electric coupling can find a natural playground in small-gap insulators, such as chalcogenides, and can bring brand new assets into the field of electrically-controlled semiconductor spintronics.

## 1. Introduction: spin orbit coupling as key ingredient in modern materials science

In recent years, there have been increasing interests in phenomena emerging from relativistic electrons in a solid. The spin-orbit interaction couples spin degrees of freedom with electronic orbits, therefore providing a link between the spin-space and the crystalline lattice. Generally considered as a "weak" interaction - and consequently neglected for long-time – spin-orbit coupling (SOC) is nowadays regarded as a rich source for novel physics. Among the wealth of recently discovered intriguing phenomena, SOC leads to topologically non-trivial insulating states [1], characterized by the concomitant presence of a bulk band gap and symmetry-protected conducting surface states, or it can give rise to Dresselhaus and Rashba effects (i.e. k-dependent spin-splitting in the band structure) in materials which lack inversion symmetry[2]. We recall that, according to the (*k*-linear) Rashba effect,[2] the (otherwise spin-degenerate) dispersion of free-like electrons, with momentum *k* and effective mass *m\** subject to a potential gradient (i.e. an electric field $E_z$) and in the presence of SOC, splits into two branches for oppositely spin-polarized states (labeled with + or -), with energies:

$$E_{\pm}(k) = \frac{\hbar^2 k^2}{2m^*} \pm \alpha_R k$$

where the so called *Rashba parameter* $\alpha_R$ (proportional to $\lambda E_z$, $\lambda$ being the spin-orbit constant) represents the strength of the Rashba effect. The *Rashba momentum* $k_R = |m^*| \alpha_R / \hbar^2$ quantifies the mutual shift of the split bands. As for Rashba-like effects, the research has so far mostly focused on materials surfaces or interfaces (i.e. where the inversion symmetry is intrinsically broken), because of their potential applications in the field of semiconductor spintronics, aiming at an all-electric control of spin transport in novel devices. Furthermore, it has been recently noticed that topological insulators and noncentrosymmetric materials displaying Rashba effects share some similarities, due to the common key ingredient, spin-orbit coupling. In this framework, a major breakthrough was provided by BiTeI, the first known noncentrosymmetric semiconductor showing huge Rashba-like spin-splitting in its *bulk* (not surface) band-structure, [3,4] driven by the presence of a nontrivial ordering of the bands near the Fermi surface[5]. As a matter of fact, BiTeI has been predicted to undergo a topological transition under pressure[6], pointing towards possible connections between topology and Rashba physics in the same (class of) materials.

## 2. Coexistence and coupling between "Bulk" Rashba effect and ferroelectricity in GeTe.

An interesting class of bulk materials lacking inversion symmetry – a prerequisite for Rashba effects to arise - is represented by ferroelectrics, i.e. materials displaying, below a certain critical temperature, a long-range dipolar order with a permanent ferroelectric polarization switchable by an electric field. Novel functionalities could be exploited in semiconductor spintronics in the presence of the coexistence and possible coupling between ferroelectricity and SOC-induced Rashba effects. Indeed, the potential relevance of this phenomenology has just started to be discussed in our recent theoretical work,[7] where the reversal of ferroelectric polarization in GeTe was theoretically predicted to cause a full reversal of the spin texture (i.e. of the Rashba parameter) and where a Datta-Das spin-transistor architecture was proposed in order to exploit this peculiar property. We here recall our main theoretical findings and address the reader to our original publication [7] for full details.

GeTe is probably the "simplest" known ferroelectric[8]: below a Curie ordering temperature of about 720 K, it shows Ge and Te ions displaced from their ideal rocksalt sites along the [111] direction, as well as a rhombohedral distortion (space group *R3m*, No. 160). We remark that the type of ferroelectricity in GeTe is strongly debated between a displacive-like and an order-disorder-transition.[9] For our specific purposes, this issue is, however, not so crucial, as we consider the situation below the Curie temperature with ideally-perfect collective FE order. We performed first principles calculations via density functional theory (DFT), within the Generalized Gradient Approximation (GGA-PBE) [10], including spin-orbit-coupling and using the VASP code[11], for ferroelectric GeTe. We show the calculated band structure in Fig.1 a), b): a spin-splitting occurs for electrons/holes with k-vectors belonging to the Brillouin zone plane perpendicular to the direction of ferroelectric polarization (i.e. around the Z point and specifically along the Z-A and Z-U lines). This Rashba splitting is indeed anomalously large for holes in the valence band maximum, opening the way to its exploitation in spintronic devices operating at room temperature.

As shown in Fig. 1 c) and d), the switch of the ferroelectric state - obtained by switching the Ge-Te relative displacements – implies that the spin-expectation values – both in plane and out-of-plane [cfr. Fig. 1) f)] – correspondingly switch their direction. Crucial for applications is therefore that switching the ferroelectric polarization causes a full reversal of the spin texture, as the spin direction in each sub-band changes by 180 degrees upon reversal of polarization, thereby allowing its electrical control. Strong hexagonal "warping" effects [cfr. Fig. 1) e)] as well as the presence of a significant out-of-plane spin-polarization are consistent with the deviation of the results from a linear Rashba model (according to a simple parabolic band-structure) and can be interpreted by including cubic terms in k in the Rashba Hamiltonian.[7]

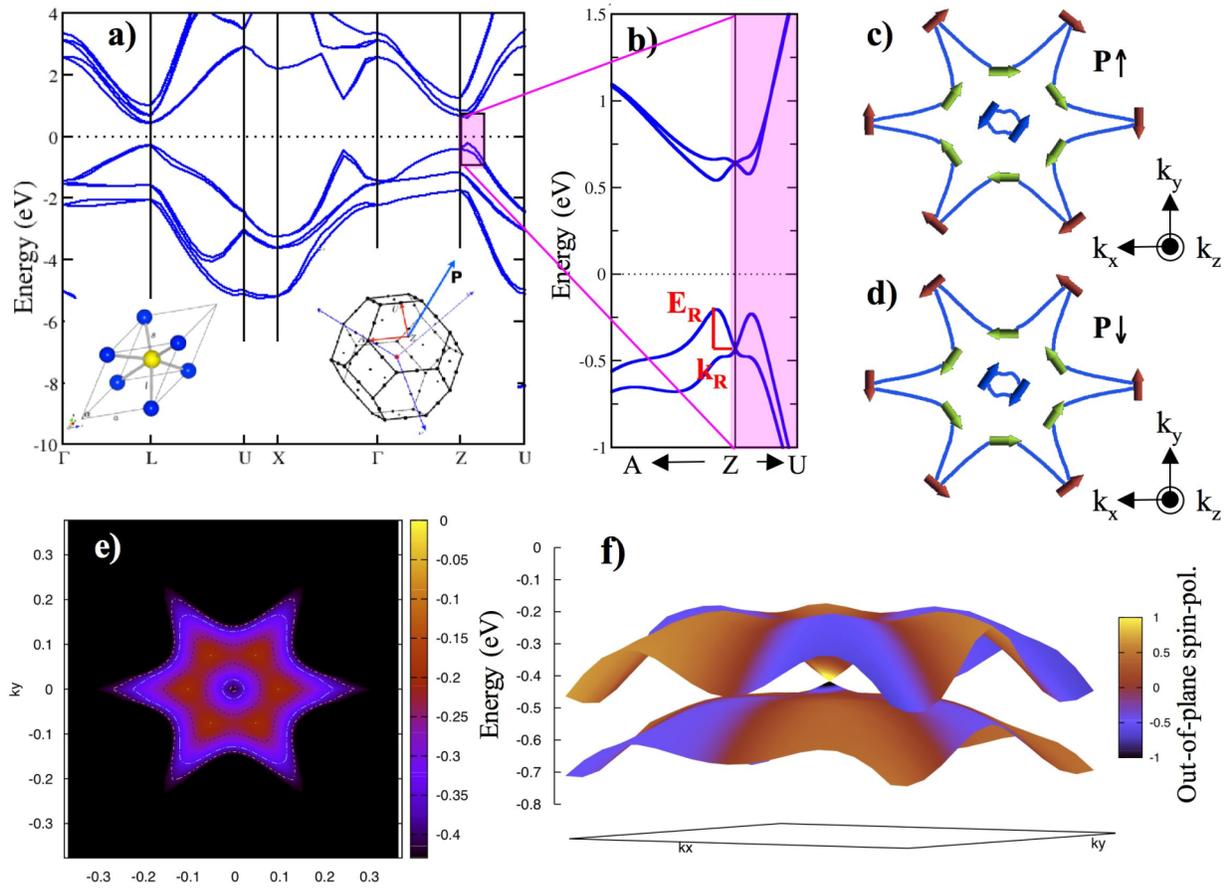

*Figure 1.* DFT results for ferroelectric GeTe. a) Band structure along main symmetry lines in the rhombohedral R3m Brillouin zone. The left (right) inset shows the distorted crystal structure (Brillouin zone, BZ), with polarization pointing along the [111] pseudocubic direction; b) Zoom of the band structure in the BZ plane A-Z-U perpendicular to polarization P (i.e. where Rashba effects show up) for bands close to the Fermi level. The Rashba parameters $E_R$ and $k_R$ are highlighted. c) Isoenergy cuts at an energy of -0.47 eV around Z in the A-Z-U plane. The spin-expectation values for holes are shown by arrows for polarization direction parallel to [111]; d) same as panel c) but for opposite polarization direction: the spin texture is reversed. e) Isoenergy cuts in the BZ plane perpendicular to P for energies ranging from -0.4 eV to 0 eV (see color scale on the right side of the panel); f) Band plots of the two upper occupied valence bands as a function of k-vectors around Z, along with the out-of-plane spin-polarization (see color scale on the right side of the panel).

As previously mentioned, the strength of the Rashba effect in GeTe can be quantified through several parameters: the k-space shift $k_R$, the Rashba energy $E_R$ and the "Rashba parameter" $\alpha_R$. By neglecting the third order terms in the nearly-free-electron approximation[7], the Rashba parameter can be related to $E_R$ and $k_R$ by $\alpha_R = 2 E_R / k_R$. The parameters are estimated for Rashba-splitted bands along the Z-A direction in k-space, in which the spin-splittings are larger. $k_R$ is evaluated as the Rashba-induced momentum offset of the valence band maximum (VBM) or conduction band minimum (CBM) with respect to the high-symmetry point, Z. $E_R$ is calculated as the difference between the VBM or CBM estimated at $k_R$ and the corresponding energy values at the Z point. For GeTe, we calculated $k_R = 0.094$ Å$^{-1}$, $E_R$ is estimated as 227 meV and 120 meV (in turn leading to $α_R$ equal to 4.9 eVÅ and 2.5 eVÅ**)** for VBM and CBM respectively. We remark that the choice of the DFT exchange-correlation functional does not seem to strongly affect the Rashba-related quantitites:[7] for example, when using accurate non-local hybrid-functionals according to the Heyd-Scuseria-Erzenhof (HSE) approach[12], one obtains for the VBM $E_R = 187$ meV, $k_R = 0.088$ Å$^{-1}$ and $α_R = 4.2$ eVÅ (although the energy band gap at the Z-point increases from 0.74 eV in GGA to 0.96 in HSE), so that variations with respect to GGA can be quantified on the order of 10%.

In order to show a comparison with other Rashba-like systems, we report in Table 1 the relevant parameters for different materials: our values show that the *bulk* Rashba effect in GeTe is indeed gigantic. In particular, we briefly discuss BiTeI, with a (noncentrosymmetric) hexagonal crystal structure of BiTeI (*P3m1*) consisting of alternating layers of Bi, Te, and I ions. On the basis of angular-resolved photoemission spectroscopy (ARPES),[3,4] magnetotransport, optical spectroscopy,[20] as well as density-functional calculations (including SOC),[3,5] the BiTeI polar structure and the large spin-orbit interaction give rise to a giant Rashba-like spin splitting in the bulk conduction and valence bands of BiTeI, also reflected in a giant and robust spin-splitting at its surface. Similar results were also found for related compounds, such as BiTeBr.[15,16] However, the presence of bistability granted by GeTe ferroelectricity represents a remarkable improvement over the Rashba phenomenology observed in BiTeI and BiTeBr. The power of accurate first-principles simulations joint to modelling capabilities based on symmetry-considerations[7] is therefore evident in this field and constitute grounds for a deep understanding of ferroelectrics with strong Rashba effect.

## 3. Ferroelectric Rashba Semiconductors (FERSC)

### 3.1 Definition and possible technological applications

Following the results on GeTe remarkable properties, we coined the acronym "FERSC", standing for Ferro-Electric Rashba Semi-Conductors, a class of materials which we consider as an integration of different subfields: i) *Ferroelectricity* (granting the switchability of ferroelectric polarization by an electric field, traditionally exploited in non-volatile memory elements), ii) *Rashba effects* (bringing in spin-degrees of freedom and usually exploited in logic semiconducting-based architectures) and iii) *Semiconductor Spintronics* (granting the integration with existing semiconductor-based technology). These concepts are pictorially summarized in Fig.2. We further remark that, as already pointed out, one of the most interesting features of FERSC is the link between spin texture and ferroelectric polarization. FERSC therefore also branch into the wide and rich field of magnetoelectricity, with intriguing future perspectives given by the effect that external magnetic and electric fields might have on a material where spin and dipolar degrees of freedom are already active and strongly intertwined. Although the focus of the present review is to present the novel microscopic mechanism brought by the interplay between ferroelectricity and Rashba effects and, as such, has its main relevance from the basic science point of view, we here briefly mention that the FERSC GeTe properties might offer intriguing perspectives also from the technological point of view. In fact, within a FERSC, the Rashba induced spin precession of an injected current is controlled by its ferroelectric (FE) state. As such, when in contact with normal magnets, a FERSC can for example be used as non-volatile channel in a spin-FET device with two additional magnetic elements, working as spin injector and detector. The spins within the ferromagnetic elements are then coupled through the current and determine the resistance. Thus the modified "Datta-Das" spin-FET architecture, as it was originally proposed in Ref. [7], has three memory elements (two magnets and one ferroelectric) and one logic channel operating with spins and controlled by the FE polarization, thereby allowing to *combine logic and storage functionalities.*

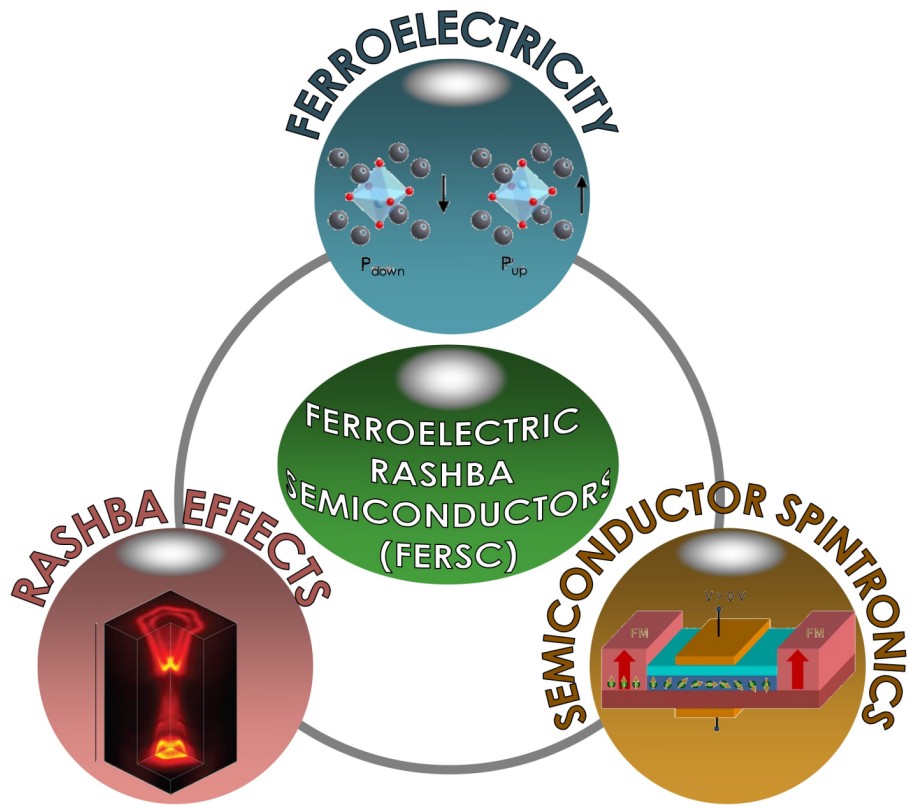

*Figure 2. FERSC, pictorially represented as the merging and coupling between different materials-science and technology subfields.*

Table 1. Rashba parameters (energy splitting $E_R$, momentum offset $k_R$ and Rashba parameter $\alpha_R$) for several systems.

| System | $k_R$ (Å$^{-1}$) | $E_R$ (meV) | $\alpha_R$ (eVÅ) | Ref. |
|---|---|---|---|---|
| Au(111) (Surface) | 0.012 | 2.1 | 0.33 | Ref.[13] |
| InGaAs/InAlAs (Semiconductor interface) | 0.028 | <1 | 0.07 | Ref.[14] |
| BiTeI (bulk, polar) | 0.052 | 100 | 3.8 | Ref.[3] |
| BiTeBr (bulk, polar) | ~0.05 | ~50 | ~2 | Ref.[15,16] |
| LaAlO3/SrTiO3 (Oxide interface) | ? | <5 | 1-5 10$^{-2}$ | Refs.[17,18,19] |
| GeTe (bulk, ferroelectric) | 0.09 | 227 | 4.8 | Ref.[7] |

## 3.2 Materials: GeTe and beyond

So far only one ferroelectric material – GeTe – has been predicted, [7] to our knowledge, to show a giant Rashba spin-splitting. Our theoretical predictions clearly call for an experimental confirmation. Efforts will have to be devoted to show the presence of Rashba spin-split bands via Spin-resolved Angle-Resolved Photoemission Spectroscopy (ARPES) and, later on, to demonstrate that Rashba spin-splitting and ferroelectricity are intimately linked. GeTe itself, however, has several pitfalls from the experimental point of view. In fact, whereas ferroelectric displacements have been clearly observed with many experimental techniques, [21] GeTe shows a high tendency to form Ge vacancies[22,23]. This in turn leads to a p-degenerate semiconducting behavior, calling into question the possibility to switch the ferroelectric state in such a "conducting" material, therefore hindering the control of spin-texture via an electric field. In this same framework, a comment is in order: recently, it was theoretically shown [24] for BaTiO3 that ferroelectric displacements develop up to a critical concentration (0.11 electron per unit cell volume), in agreement with experimental data. This result reveals that ferroelectricity and conductivity can indeed coexist and we expect a similar situation to occur in GeTe as well.

Given some disadvantages of Germanium Telluride, the main future objective from a materials design point of view will therefore be to expand the class of FERSC materials beyond GeTe in the aim of identifying a "strong" (i.e. not-leaky) ferroelectric, where the full-reversal of the spin-texture via an electric field switching is predicted and offered to experiments for confirmation. This however implies a deep understanding of which are the necessary conditions – in addition to (obvious) ferroelectricity and large SOC - that lead to a giant Rashba spin-splitting. Suggestions came from Ref.[5], where it was pointed out that one of the conditions necessary for a large Rashba splitting is to have the same orbital character for the cross-gap states mixed upon SOC, a requirement not easily met in usual ferroelectrics (such as oxides). At the same time, one will have to increase the band gap with respect to GeTe (with an estimated gap [7] of the order of half an eV), so as to suppress conductivity. Among possible semiconductor ferroelectrics, binary or ternary IV-VI chalcogenides represent a rather simple but instructive playground, where different properties of relevance for Rashba effects (k-dependent spin splitting around high-symmetry points, Rashba parameter, etc) may be analysed by means of first-principle tools. The effect of strain or pressure on GeTe properties might also be examined, along with doping/alloying (for example with C, Pb, or other dopants). The materials design of a successful FERSC should lead to a material that i) keeps the ferroelectric behaviour, i.e. the total energy should display the typical double-well potential as a function of the ferroelectric distortion and the theoretical estimate of ferroelectric polarization should be of the order of $\mu C/cm^2$ and ii) has a high resistivity (or correspondingly, the tendency to form vacancies should be reduced or possibly suppressed, compared to GeTe).

When moving away from IV-VI semiconductors, one can in principle explore several routes, looking at known ferroelectric material classes and/or including systematic crystallographic database mining for polar systems with "heavy" elements (to achieve a large SOC) and "small" gap (for an efficient coupling through SOC). Compared to (traditional) ferroelectric oxides, the substitution of oxygen with S, Se, Te seems promising, due to the lower bond ionicity of chalcogenides. For example, semiconducting chalcogenides such as $M_2P_2X_6$ ($M=Sn^{2+},Pb^{2+}$ and $X=S,Se$) are ferroelectric due to the Sn or Pb stereochemically active lone-pair. In parallel, high-throughput approaches,[25] based on large-scale DFT calculations performed for real and hypothetical systems, could be adopted for FERSC. A similar approach led for example to the identification of largely unknown classes of semiconducting ferroelectric materials, such as $P6_3mc$ "LiGaGe-type" systems. In a recent work by Bennett et al.[26], several potential candidates satisfying the identified prerequisites for giant Rashba splitting are present (based on Bi or Sb or Sn), including some ferroelectric semiconductors with a direct band-gap. The latter condition might be of relevance even for novel Rashba-based magneto-optical devices (recall that GeTe has an indirect band-gap, hindering its use for any optical device based on the giant Rashba spin splitting).

## 4. Conclusions

In summary, we have proposed a brand new spin-electric coupling mechanism deriving from the interplay between spin-orbit-driven Rashba effect and ferroelectricity: in GeTe, the prototype of Ferroelectric Rashba Semiconductors (FERSC), the spin-texture is controlled and switched via an electric field. As such, FERSC therefore represent breakthroughs in different aspects of science and technology: i) from the fundamental point of view, FERSC offer playground for new and completely unexplored microscopic understanding of "cross-coupled" spin-electric effects; ii) from the materials science point of view, FERSC represent a novel class of multifunctional materials, and, as such, there is room for their development, accompanied by a deeper understanding and optimization; iii) from the technological point of view, the FERSC phenomenology might be exploited in prototypes of an entirely new generation of semiconductor spintronic devices, with a remarkable integration of logic functionalities and non-volatility.

## 5. Acknowledgement

I would like to acknowledge Dr. P. Barone and Dr. D. Di Sante (CNR-SPIN, L'Aquila) and Prof. R. Bertacco (Politecnico di Milano) for many fruitful discussions. Computational support from PRACE supercomputing grant "TRASFER" at Barcelona Supercomputing Center is gratefully acknowledged.